# XML Query Processing and Query Languages: A Survey

Mikael Fernandus Simalango
Graduate School of Information and Communication
Department of Computer Engineering, Ajou University, South Korea
Email: mikael@ajou.ac.kr

Abstract-*Today's database is associated with interoperability between different domains and applications. This consequently results in the importance of data portability in database. XML format fits the requirements and it has been increasingly used for serving applications across different domains and purposes. However, querying XML document effectively and efficiently is still a challenging issue. This paper discusses query processing issues on XML and reviews proposed solutions for querying XML databases by various authors.*

## I. INTRODUCTION

XML[10] serves dual functionalities as markup language and data format. It separates presentation and data thus offering independency and flexibility for content association. Due to this nature of flexibility, data interchanged between two very different systems can use XML as the data format. XML tree-like structure is intuitive, human readable, and easy to understand. With the help of XML schema or DTD, the type and attributes of each tag usable for certain XML document can be well defined.

An XML query language defines more comprehensible and structurized construct for conducting operation on an XML document or various XML documents. For processing the query, an XML query engine or processor translates the syntaxes and executing the operations hinted by the query. Output is returned after process and processing time is projected to be minimum thus alluding efficient processing.

However, as non binary format, performing query over XML data pertaining to arbitrary applications is still an intriguing issue. In the past, most effort was put to design query processor to support declaration in query languages. These days, the issue has shifted to relational XML storage and integration with data management system.

The rest of this paper is arranged as follows. We initially review the evolving path of XML query languages. Then, we provide different approaches for xml query processing by extracting the ideas and comparing the proposals. Finally, we provide possible direction for future xml database and sum up our conclusion.

## II. XML QUERY LANGUAGES

Since 2007, XQuery[9] which is an extension of XPath[10] has been recommended by W3C as query language for XML document. However prior to the establishment of W3C standard, there had been several researches proposing query languages for XML.

D. Maier elaborated desired characteristics of XML query language[12]. His criteria were massively used as reference for development of some XML query languages. Important criteria in his proposal include xml output of the query, independence of schema, schema exploitation if possible, and optimized query operations. The operations defined in the proposal are selection which is choosing document or document element, extraction which is pulling out elements of a document, reduction that is realized as removing sub-elements, restructuring or constructing a new set of element instances, and combination as merging operation carried out over two or more elements resulting in only single element.

| Query Langs. | Lang. Type | Input model | Class of query | Public recognition |
|---|---|---|---|---|
| XML-QL | functional | XML | Pattern matching | 1998 |
| Lorel | declarative | OEM | Path expressions within OQL | 1997 |
| Quilt | functional | XML | Quilt expressions | 2000 |
| XQL | functional | XML | XQL based on path expressions | 1999 |
| XQuery | functional | XML | XQuery | 2007 (recommendation) |

Table 1 XML query languages in comparison

XML-QL[14] is an XML query language which provides support for querying, constructing, transforming and integrating XML data. This language reflects XML as semistructured data that have irregular or rapidly evolving structure. XML-QL uses element patterns to match data in an XML document. An extension of XML-QL named Elixir[15] was proposed to support ranked queries based on textual similarity.
Pros: schema aware, nested queries
Cons: heavily pattern based, a priori knowledge of data structure is usually required, cumbersome syntax.

Lorel[16] is early query language for semistructured data. It uses OEM (Object Exchange Model) as the data model for semistructured data. For querying the elements, Lorel extends OQL (Object Query Language) by relying on coercion at a number of levels to restrain the strong typing of OQL. Lorel also extends OQL with path expressions so that user can specify patterns that are matched to actual paths in referred data.
Pros: easy syntax
Cons: dependant on OQL parser, limited functionalities

Quilt[13] is a functional language in which a query is represented as expression. There are seven principal forms of Quilt expressions which are path expressions, element constructors, FLWR expressions, expressions with operator and functions, conditional expressions, quantifiers, and variable bindings. Besides join operations, quilt also support nested expressions hence it basically support subquery within a single query. Significant features of Quilts were used for the development of XQuery.
Pros: robust functionalities, subqueries
Cons: no support for textual similarity

XQL[17] uses path expressions hence its basic constructs correspond directly to the basic structures of XML. Due to this nature, XQL is closely related to XPath. In XQL, document nodes play a central role. Nodes have identity and they retain their identity, containment relationships, and sequence in query results. The nodes themselves may come from variety of different sources. However, XQL does not specify how these nodes are brought to the query. XQL also supports joins and some functions.
Pros: shorter expressions
Cons: semantics may not be very intuitive

XQuery[9] had been a moving target for some time before it was established as W3C recommendation in 2007. A big part of XQuery semantics adopts Quilt's. XQuery uses XPath for path expressions and FLWOR structure for describing the whole query. As a recommended standard, a lot of researches nowadays discuss the method of optimizing XQuery translation and processing by a query processor and integrating XQuery into a full-fledged XML database management system.
Pros: clear semantics, integration with XPath
Cons: intersection with XSL

Important characteristics of various XML query languages can be seen in Table 1.

### III. APPROACHES FOR XML QUERY PROCESSING

A query processor extracts the high level abstraction of declarative query and its procedural evaluation into a set of low-level operations[18]. Analogous to SQL processor, SQL query is translated at logical access model and then the logical access prior to accessing and returning the physical storage model. Levels of abstraction in XML query processing in comparison with SQL abstraction levels are depicted in Table 2.

| Level of Abstraction | XDBS | RDBS |
|---|---|---|
| Language model | XQuery | SQL |
| Logical access model | XML query algebra | Relational algebra |
| Physical access model | Physical XML query algebra | Phyiscal DB-operators |
| Storage model | XTC, natix, shredded documents, etc | Record-oriented DB-interface |

Table 2 XDBS vs. RDBS abstraction levels

From Table 2, XDBS denotes XML database management system and RDBS are Relational Database Management System. The language model is designed to meet the demands of [12] which are reflected in the language ability to perform search functionality and document-order awareness hence document-centric characteristics and later on the data-centric characteristics which is associated with powerful selection and transformation. The semantic processing should then be able to analyze the query and transform it into an international representation to be used throughout subsequent optimization steps.

Logical access model should implement algebraic and non-algebraic procedure to optimize the internal representation of the query. Non-algebraic optimization minimizes intermediary results by restructuring the query and executing most selective operations as early as possible. Algebraic optimization will transform the internal expression into a more optimized expression in a semantics-preserving manner.

Physical access model is related to system-specific issue. At this level, each logical algebra operator will be decomposed into corresponding physical operators. The goal of this step of optimization is a query executing plan (QEP) which is arranged of chosen physical operators and their sequences of execution.

Finally, the storage model affects the rate of QEP. For optimized query processing, appropriate storage model should be deployed in order to minimize I/O costs, CPU costs, storage costs for intermediary results, and communication costs. Currently used storage models comprise LOBs (Large Objects), certain XML-to-relational mappings (shredded documents), or native storage formats like Niagara[19] and Timber[20]. The relational XML data model and native storage model attract more attentions indicated by various proposals for respective overlying query processors.

Various XML query processors have been proposed for more optimized query processing. Referring to the abstraction levels, we'll divide the query processors into three categories based on their storage models: flat-file processing, relational processing and native storage processing.

**Query Processing on Flat File Scheme**

In flat file processing, for example when XML is saved as LOBs, query is executed after all XML data is loaded and scanned by the query processor. This surely results in poor performance when the size of file is big and temporary storage in memory is not feasible. However, some algorithms were authored to improve the query processing. N. Bruno *et al*[21] studied different techniques for processing XML queries: y-filter, index-filter, and pathstack. Y-filter is query processing by augmenting prefix tree representation of input queries as an NFA (Non-deterministic Finite Automaton) which will output all matches of the queries. The index filter technique uses indexes built over certain tags of the input XML document. PathStack which is a series of linked stacks is later created for each query node in a path query in order to track the data nodes. Figure 1 shows how indexes for an XML document are created using this approach.

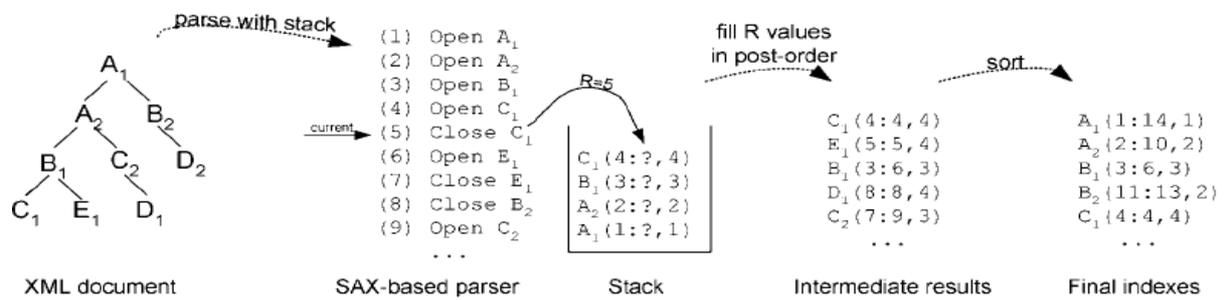

Figure 1 Index creation in index-filter technique

**Query Processing on Relational Structure**

In this approach, XML document or information related to XML document is stored in relational database. This step is taken because relational database performs better indexing than simple index creation like in previous approach. RDBMS engine will instead perform the query processing by translating XQuery into SQL, running the SQL query and serialize the XML result.

Relational storage schemes for XML documents can be classified into three groups: no XML schema scheme, based on XML schema, and user defined. In case there is no schema provided, relational schema should be derived from the data. After schema exists relational schema will be created which contains relationship among root element and all sub-elements.

In [2], the authors divides relational scheme into scheme-oblivious and scheme-conscious approach. Scheme-oblivious approach maintains a fixed schema by caputing the tree structure of XML documents. In contrast, scheme-conscious approach creates a relational schema based on DTD/schema of the XML first and based on the schema, primary-key foreign-key joins in relational database are set up to model parent-child relationships in the XML tree. The authors built SUCXENT++ and observed that schema-oblivious approach could also outperform schema-conscious approach.

The authors in [2] also provided comparisons for other different schemes like EDGE, XRel, and XParent which are not discussed here for brevity.

BEA/XQRL[4] is a query processor that implements relational scheme using XQuery. Query is parsed and optimized by query compiler. For eliciting the query, XDBC interface functions as an interface between frontend application and query processor. The compiler will then generate a query plan to optimize the query. XML data is represented as stream and parsed as input by the XML parser. Runtime operators containing function and operator libraries will process the stream and provide output based on the query plan.

Figure 2 depicts the overview of BEA streaming XQuery engine.

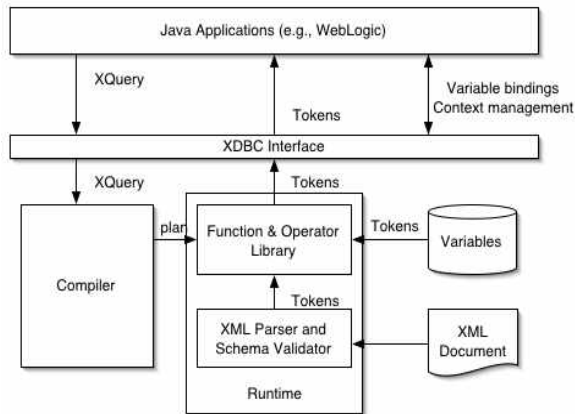

Figure 2 Overview of BEA

MonetDB[5] is another query processor for XQuery which is constituted by Pathfinder compiler on top of MonetDB RDBMS. It also has XQuery runtime module that utilizes loop-lifted staircase join (a method for evaluating XPath location in a single sequential scan) as a physical operator so that the query processing can be improved.

**Query Processing on Native Storage Scheme**

Using this approach, XML elements are assigned label. The purpose of the labeling is to create unique identifiers that will be useful for query processing. There are many labeling schemes which take into account trade-off between space occupancy, information contents, and suitability to updates. The most frequently used is region-based labeling scheme. The idea of this scheme is to label elements to reflect nesting. Figure 3 shows the labeling scheme for simple nesting. The final label denotes (start, end, level) status for the node.

```
<a>Unnested content and <b>nested one</b></a>
1. Label <a> with [1,2] and b with [2,1]
2. Add nesting level
     Label <a> with [1,2,1]
     Label <b> with [2,1,2]
```

Figure 3 Region-based labeling scheme

Another labeling scheme is ORDPATHS which is implemented in MS SQL server. This scheme labels each node by a sequence of integer numbers. Order, depth, parent, and ancestor-descendant relationships are recorded in this scheme.

The XML document will later stored as persistent trees. If disk is used as storage means, XML nodes will be split among disk page. Node representation is optimized based on fixed page size.

Efficient query processing in native storage is achieved by stack-based algorithms like StackTreeDesc[22] and holistic twig joins[23]. StackTreeDesc algorithm uses stack structure to cache parent elements' label and when path to destination child node is reached, information from stack is combined with child label and returned as results in descendant order. Subsequently, stack is emptied for the next operation. On the other hand, holistic twig joins tries to avoid constructing intermediary results when matching twig (search for predicate or label) patterns.

NaxDB[7] uses native approach and supports XQuery and XUpdate processing. In NaxDB, hierarchical tree of linked objects from XML data is stored using object oriented extensions of MaxDB from MySQL. MaxDB system architecture are built on top of three subsystems: a database client that enables users to write queries and receive results, a database server which is the core subsystem, and persistent object manager which is responsible for persistently storing XML data.

### IV. TOWARD FUTURE XML DATABASE MANAGEMENT SYSTEMS

Future database management system is associated with application mash-up and versatility. It will operate across different platforms thus it has to handle interoperability among data. Data can be static or in a form of stream and its flow may vary from low-density stream to high-density stream. Database management system, should be aware of those characteristics and be able to perform well by minimizing the costs.

This paper has reviewed progress toward XML database management system. Current trends inclined to relational scheme where query for XML data is translated into declarative SQL to speed up the indexing process and node solicitation.

Future researches can be targeted to design better pathfinding algorithm like in [3,6], alternative query processing like in [1], and support for transactional XML databases.

## V. CONCLUSION

Since XQuery is now a de facto standard for query language over XML, nowadays a lot of effort is put to achieve more efficient and optimized XML query processing. Current trends are inclined to relational scheme which consolidates XML with features of RDBMS. However, several challenges for the realization of scalable XML database management system still exist and future researches should address them pretty well.